\documentclass[11pt,twoside]{article}

\usepackage{asp2006}
\usepackage{epsf}
\usepackage{lscape}
\usepackage{graphicx}
\begin{document}
\title{Inhomogeneous Galactic halo: a possible explanation
	for the spread observed in s- and r- process elements}   
\author{Gabriele Cescutti}   
\affil{Dipartimento di Astronomia, Universit\'{a} di Trieste, via
	G. B. Tiepolo 11, 34131 Trieste, Italy}    

\begin{abstract} 
The considerable scatter of the s- and r-process elements observed in
low-metallicity stars, compared to the small star to star
scatter observed for the alpha elements, is an open question for the chemical evolution 
studies.
We have developed a stochastic chemical evolution model, in which the
main assumption is a random formation of new stars, subject to the
condition that the cumulative mass distribution follows a given 
initial mass function.
With our model we are able to reproduce the different features of $\alpha$-elements 
and s-and r-process elements.
 The reason for this resides in the random birth of stellar masses coupled with 
the different stellar mass ranges from where  $\alpha$-elements 
and s-and r-process elements originate.
In particular, the sites of production of the alpha elements are the whole range of 
the massive stars, whereas the mass range of production for the s- and r-process
elements has an upper limit of 30 solar masses.

\end{abstract}

 
\subsection*{Inhomogeneous chemical evolution model for the Milky Way halo}  

We model the chemical evolution of the halo of the Milky Way 
for 1 Gyr. We consider the halo has formed by means of the assembly of 
many independent regions with 	a typical volume of $10^{6}pc^{3}$.
 Each region does not interact with  the
others. Inside each region the mixing is assumed instantaneous.
In each regions we assume an infall episode with a timescale of 1Gyr and a threshold 
in the gas density for the star formation.

When the threshold density is reached, the mass of gas which is
 transformed at each timestep in stars, $M_{stars}^{new}$, 
is proportional to $\rho_{gas}^{1.5}$.
We simulate the birth of stars until the sum of the stars exceed the $M_{stars}^{new}$.
The mass of each star is assigned with a random function in the range 
0.1 and 80 $M_{\odot}$, weighted according to the IMF of Scalo (1986).
In this way, in each region, for each timestep, the $M_{stars}^{new}$ is the same but 
the total number and the masses of the stars are different.
The model follows the chemical evolution of more than 20 elements
in each regions.
The model  parameters of the chemical evolution (SFR, IMF, 
stellar lifetime, nucleosynthesis, threshold), have been taken from
the homogeneous model by Chiappini et al.(1997).
Our model shows the spread that is produced by different nucleosynthesis sites 
on the chemical enrichment at low metallicity, where the number of stars is low and
the random effects in the birth of stellar masses can be important.


\subsection*{Results}

In the figure we show our simulation (blue dots)
compared to the observational data (red open triangles). 
The data are by Cayrel et al.(2004) 
and other authors.
The black line is the prediction of the homogeneous model. 
Concerning neutron capture elements we show Eu 
and for  $\alpha$-elements Si.
\begin{figure}
\plottwo{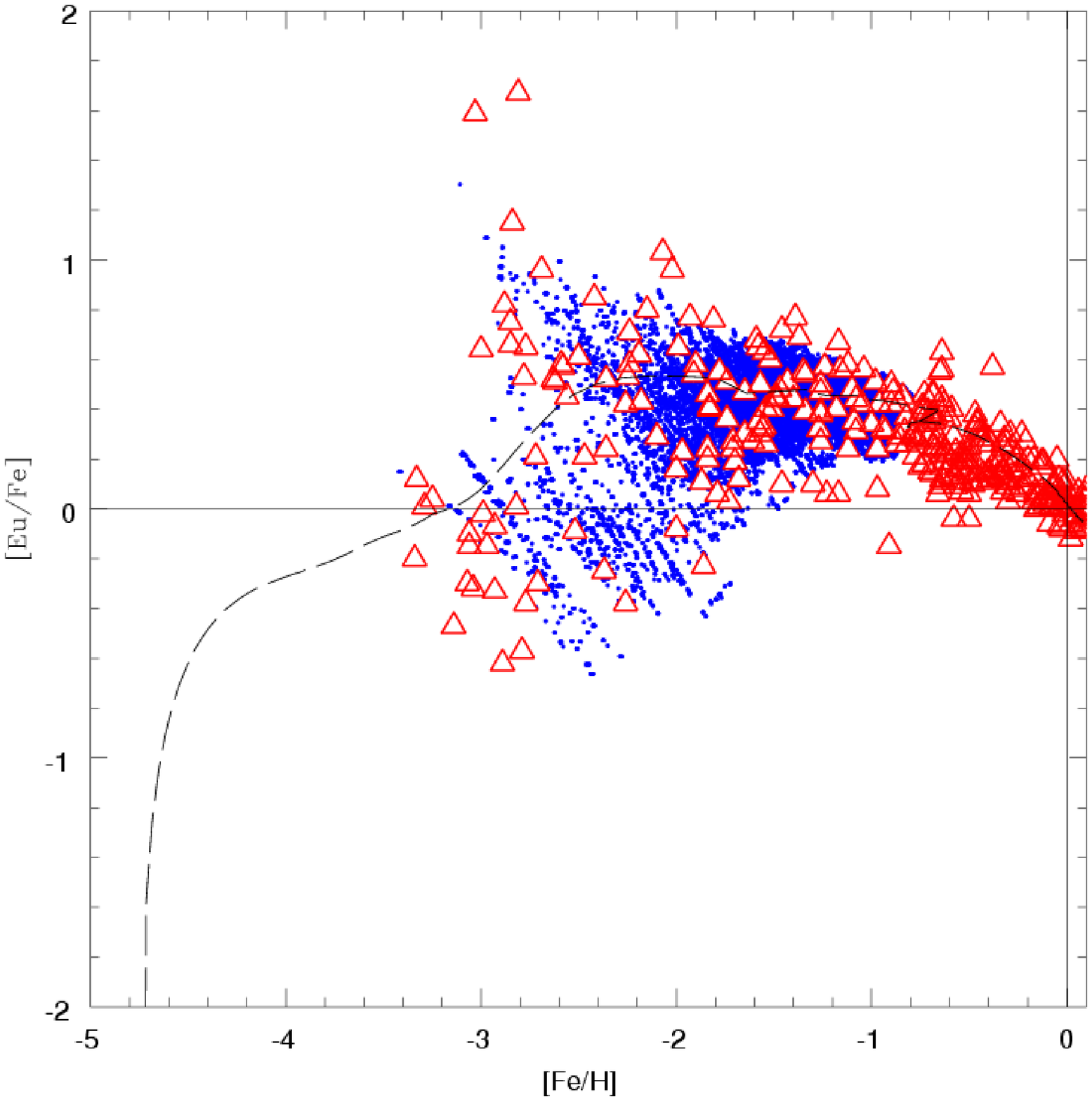}{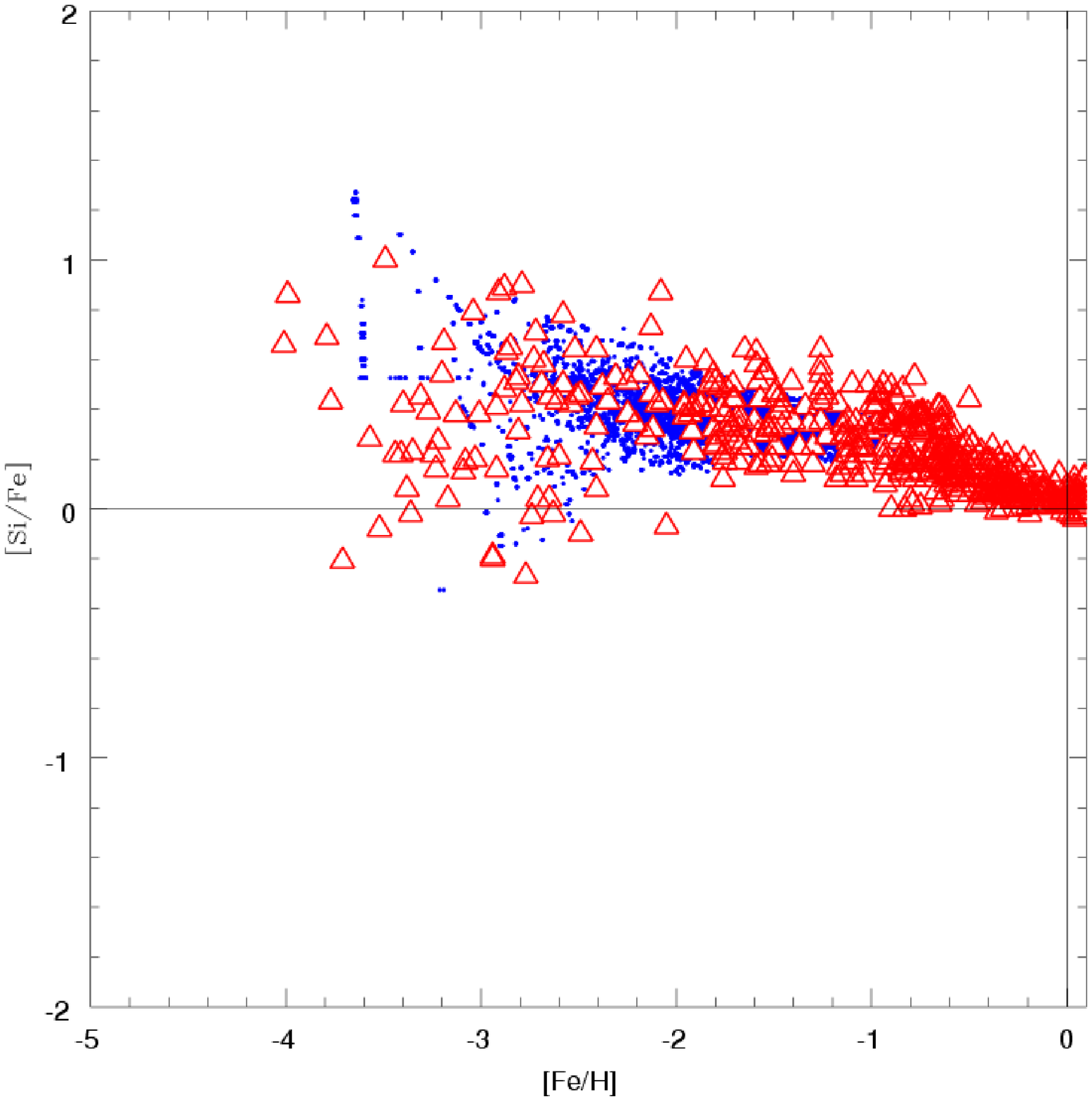}
\caption {[El/Fe] vs [Fe/H]. The abundances of simulated stars in blue dots,
	observational data in red open triangles. The black line is the prediction of the homogeneous
	model}
\end{figure}

The enrichment of metal poor stars is due to the massive stars.
The sites of production in massive stars for the
 $\alpha$-elements  and neutron capture elements are
 different for the nucleosynthesis prescriptions we adopted:
the $\alpha$ elements and Fe are produced in the whole range of massive stars
(see Fran\c cois et al 2004);
the neutron capture are produced only up to 30 $M_{\odot}$ (see Cescutti et al. 2006).
In regions with many stars less massive than 30 $M_{\odot}$, the
 ratio of neutron capture elements over Fe is
high. The opposite happens in regions where the most of the stars are more massive
than 30 $M_{\odot}$. This fact produces in our inhomogeneous model a large spread for
the neutron capture elements.
This does not happen for $\alpha$-elements which are produced in the same range 
of Fe.

\section*{Conclusion}
A random birth of stellar masses, coupled with 
the different mass range  responsible for the production of
 $\alpha$-elements	and neutron capture elements respectively,
 can explain the large spread in the abundances of 
metal poor stars for neutron capture elements and the smaller
spread for $\alpha$-elements. 



\acknowledgements The author warmly thanks Francesca Matteucci, 
Simone Recchi and Antonio Pipino for many enlightening discussions.




\begin{thebibliography}{}
\bibitem[]{} Cayrel R., Depagne E.,Spite M. et al., 2004, A\&A, 416, 1117 			   
\bibitem[]{} Cescutti G., Fran\c cois P., Matteucci F., Cayrel R., Spite M., 2006, A\&A, 448, 557  
\bibitem[]{} Chiappini C., Matteucci M.F., Gratton R.G.,   1997, ApJ, 477, 765			   
\bibitem[]{} Fran\c cois P., Matteucci F., Cayrel R. et al., 2004, A\&A, 421, 613		   
\bibitem[]{} Scalo J.M., 1986, FCPh, 11, 1							   
\end{thebibliography}
\end{document}